\documentclass{aa}

\usepackage{graphicx}
\usepackage{txfonts}
\usepackage{graphicx}   
\usepackage{amsmath}   
\usepackage{amsmath} 
\usepackage{float} 
\usepackage{graphicx}
\usepackage{booktabs} 
\usepackage{hyperref}
\usepackage{geometry}
\usepackage{booktabs} 
\usepackage{ifthen}
\usepackage{lipsum}
\usepackage{makecell}
\usepackage{subcaption}
\usepackage{xspace} 
\usepackage{multirow}
\usepackage{multicol}

\usepackage[normalem]{ulem}
\usepackage[dvipsnames]{xcolor}

\newcommand{\eq}[2][]{\begin{align}
                          #2
\end{align}}

\newcommand{\abs}[1]{\left | #1 \right |}
\newcommand{\avg}[1]{\left \langle #1 \right \rangle}
\newcommand{\diff}{\mathrm{d}}

\newcommand{\lr}[1]{\left(#1\right)}
\newcommand{\lrc}[1]{\left[#1\right]}

\newcommand{\BH}{\mathrm{BH}}

\newcommand{\nbody}{\texorpdfstring{$N$-body\xspace}{N-body\xspace}}

\newcommand{\units}[1]{\,\mathrm{#1}}
\newcommand{\cmc}{\textsc{CMC}\xspace}
\newcommand{\tsunami}{\textsc{tsunami}\xspace}

\def\subinrm#1{\sb{\rm#1}}
    {\catcode`\_=13 \global\let_=\subinrm}
\mathcode`\_="8000
\def\upsubscripts{\catcode`\_=12 } 

\upsubscripts

\newcommand{\msun}{{\rm M}_\odot}

\renewcommand{\eqref}[1]{(equation~\ref{#1})}

\begin{document}

\title{Interactions among binary black holes in star clusters: \\ Eccentric gravitational wave captures and triple formation}
\titlerunning{Interactions among binary black holes in star clusters}

\author{Daniel Marín Pina
\inst{1, 2}
\and
Mark Gieles
\inst{1, 2, 3}
\and
Tomas Andrade
\inst{1, 2}
\and
Alessandro A. Trani
\inst{4}
}

\institute{
    Departament de F\'isica Qu\`antica i Astrof\'isica (FQA), Universitat de Barcelona (UB), Mart\'i i Franqu\`es 1, 08028 Barcelona, Spain\\     \email{danielmarin@icc.ub.edu}
    \and
    Institut de Ci\`encies del Cosmos (ICCUB), Universitat de Barcelona (UB), Mart\'i i Franqu\`es 1, 08028 Barcelona, Spain
    \and
    ICREA, Pg. Llu\'is Companys 23, E08010 Barcelona, Spain
    \and
    Niels Bohr International Academy, Niels Bohr Institute, Blegdamsvej 17, 2100 Copenhagen, Denmark  
}

\date{Received XX; accepted XX}

\abstract
{}
{Numerical simulations of star clusters with black holes find that there is only a single dynamically active binary black hole (BBH), at odds with the theoretical expectation of  $\sim5$ dynamically formed -- or, commonly referred to as `three-body' -- BBHs in clusters with a few hundred BHs. We test the recent suggestion that this tension is because interactions  among three-body BBHs were neglected in the theory.}
{We use the public catalogue of Cluster Monte Carlo  models to obtain a sample of strong BBH-BBH interactions, which we integrate using post-Newtonian equations of motion up to 3.5PN. We explore the nature of the BBHs involved in BBH-BBH interactions in star clusters, as well as the various outcomes: gravitational wave (GW) captures and the associated eccentricities at the frequencies of ground-based GW detectors, as well as BH triple formation and their contribution to BBH mergers via the Lidov-Kozai mechanism.}
{We find that almost all BBHs involved in BBH-BBH interactions are indeed three-body binaries and that BBH formation and disruption in BBH-BBH interactions occur at approximately the same rate, providing an explanation for the finding of a single dynamically active BBH in $N$-body models. An important implication is that the resulting rates of GW capture and triple formation are independent of uncertain initial binary properties. With the use of a population synthesis model for BBH-BBH interactions in globular clusters, we obtain a local rate of GW captures of $\mathcal{R}(z\simeq 0) \simeq 1\,\units{Gpc^{-3}}\units{yr^{-1}}$, as well as their eccentricity distribution and redshift dependence. We find that a BBH-BBH interaction is more likely to trigger a GW merger than a BH-BBH interaction. We also confirm that stable triples that are assembled in BBH-BBH interactions can merge via von~Zeipel-Lidov-Kozai oscillations, although their merger rate is lower than GW captures. Our results will help with the interpretation of future GW signals from eccentric BBHs.}
{}


\maketitle

\section{Introduction}
The detection of gravitational waves (GW) from the merger of binary black holes (BBH) by the LIGO-Virgo-Kagra (LVK) interferometers is now a routine phenomenon. More than a hundred BBH mergers have been reported in the first three observing runs \citep{O3} and the ongoing fourth observing run. Despite the rapid increase of available data, the origin of the GW sources is still unclear.

The two classical formation pathways for BBH mergers are the isolated channel and the dynamical channel. In the former, the BBH forms from a binary of massive stars in which the system evolves with negligible interaction with its surroundings and the orbit shrinks by either common envelope evolution \citep[e.g.][]{Belczynski2007, Postnov2014, Belczynski2016} or stable mass transfer \citep[e.g.][]{Inayoshi2017, GallegosGarcia2021, vanSon2022b, vanSon2022c}. Other related pathways are chemically homogeneous evolution and Population III binaries (see \citealt[Sect.~4.1]{MandelBroekgaarden2022} for a review on the topic). In the dynamical channel, the formation and evolution of the binary depend on the interactions of the BBH with other black holes (BHs) and stars in dense stellar systems. These include active galactic nuclei \citep{Bartos2017,Tagawa2020, Samsing2022,Trani2024a}, open clusters \citep{Banerjee2017, Banerjee2018a, Banerjee2018b, Rastello2019}, and globular clusters (GC) \citep{PortegiesZwart2000, Rodriguez2016, AntoniniGieles2020a, Trani2021, Antonini2023, DallAmico2024}.

A `smoking-gun' signal that differentiates the isolated and dynamical channels is the presence of residual eccentricity in the frequency band of the LVK interferometers (see also \citealt*{Antonini2024} for an analysis using the effective and precessing spin parameters $\chi_{\rm eff}$ and $\chi_{\rm p}$). Once the binary is assembled, the backreaction from GW emission very efficiently circularises the orbit, usually before the binary becomes observable at LVK frequencies \citep{Peters1964}. Therefore, for a BBH merger to have a non-negligible eccentricity, the BBH itself must be assembled very close in time to the moment of merger\footnote{For example, a binary with an initial semimajor axis of $1\units{AU}$ and initial eccentricity of $0.7$ will reach an eccentricity of $<0.1$ before its semimajor axis reaches $0.1\units{AU}$} (see e.g. Fig.~9 of \citealt{speratranireview2022}). This is only attainable in the dynamical channel, if the two BHs merge after an almost head-on approach.

In dense clusters, mergers due to the random encounter of two previously unbound BHs (single-single interaction) are rare, and the rate for this process is negligible \citep{Samsing2020}. The leading mechanisms for the production of eccentric GWs are the merger of two BHs in an interaction involving a BBH and an unbound third BH (binary-single interaction) or two BBHs (binary-binary interaction). These interactions can have chaotic intermediate states \citep{HutBahcall1983, SigurdssonPhinney1993, Trani2024b} and thus the probability of having a close encounter of two BHs is higher than in single-single interactions. 

In the models of \citet{Zevin2019} and \cite{Kremer2020}, binary-single interactions are ${\sim}10$ times more common than binary-binary interactions. However, both these channels are found to roughly contribute the same to the merger rate, as their models predict that each binary-binary interaction is more likely to trigger a merger compared to a binary-single interaction. 

It is not known whether the BBHs involved in binary-binary actions in the aforementioned simulations formed dynamically (so-called `three-body' BBHs), or formed from massive star binaries (hereafter, primordial BBHs). If they are primordial BBHs, any prediction for the GW merger rate and eccentricity distribution is sensitive to uncertainties in binary properties at birth and binary evolution, such as common envelope efficiency. However, three-body BBHs form from previously unbound BHs via dynamical interactions \citep{Heggie1975,Aarseth1976,GoodmanHut1993,Atallah2024} and can merge in subsequent interactions. In this channel, the properties of the BBHs are determined only by the cluster dynamics, and therefore are largely independent of the uncertainties in the binary stellar evolution models. Previous studies have shown that three-body BBHs dominate the production of GW mergers in clusters \citep[e.g.][]{Hong2018}. Dynamical formation tends to happen in the central, denser regions of the cluster and results in wide BBHs, most of which are easily disrupted due to interactions with single BHs. The remaining stable BBHs are likely to interact with one another, where the outcome of these BBH-BBH interactions is usually the ionisation of the wider binary. An equilibrium is achieved when a cluster only has one dynamically active BBH, as further dynamically assembled BBHs are rapidly ionised in BBH-BBH interactions \citep{MarinPinaGieles2024}. When this happens, the BBH-BBH interaction rate roughly equals the BBH formation rate.

The above process enhances the BBH-BBH interaction rate, which could lead to a higher GW capture rate than predicted by (fast) models that only consider binary-single interactions. Furthermore, according to \cite{Goodman1984} the number of dynamical binaries (which is the result of a balance between binary formation and disruption) is expected to be higher in clusters with a low number of BHs, because the central dynamics are determined by the properties of the BH sub-system \citep{BreenHeggie2013}. This points towards the higher relative importance of BBH-BBH interactions in low-mass clusters or massive clusters that have ejected most of their BHs. Indeed, different approaches for modelling BBH mergers in clusters present a discrepancy at low initial cluster masses in their ratio of in-cluster to ejected mergers. For example, \cite{Antonini2019} predicts that, for the typical GC ($\sim 10^{5-6}\units{\msun}$), nearly half of all mergers are in-cluster; whereas for the $N$-body models of lower-mass clusters of \cite{Banerjee2021b} this ratio is $\gtrsim 80\%$. A recent work \citep{Barber2024} argues that almost all mergers of non-primordial binaries happen in-cluster. GW captures in BBH-BBH interactions, which happen only in-cluster and are included in the $N$-body models but not the fast (approximate) models, may explain the differences found.

In this paper, we link the origin of BBH-BBH interactions in GCs to dynamical BBH formation and study their outcomes, including GWs with measurable eccentricity and dynamical triple formation. The paper is organised as follows: in Section \ref{sec:methods} we enumerate our initial conditions and summarise the methods used to perform the simulations. In Section \ref{sec:bbhbbh}, we present a model for the population of BBHs in star clusters and their BBH-BBH interactions. In Section \ref{sec:gwcaptures}, we put forward a model for the GW outcomes of these interactions. In Section \ref{sec:triples} we study dynamical triple formation due to BBH-BBH interactions and GW mergers due to von~Zeipel-Lidov-Kozai oscillations in these triples. Our results are discussed in Section \ref{sec:discussion}, and the conclusions are summarised in Section \ref{sec:summary}.

\section{Methods}
\label{sec:methods}

\subsection{Summary of \cmc}
\label{ssec:cmc}

We use the public suite of GC models run with \textsc{Cluster Monte Carlo} (\cmc). \cmc is a highly efficient code designed to accurately simulate the time evolution of dense collisional star clusters. It uses a H\'enon-like approach \citep{Henon1971}, where the position of stars is randomly sampled from their orbit around the cluster and the relaxation from multiple weak two-body encounters is treated as a single effective encounter \citep{Joshi2000}. This shortens the computing time and allows for an exploration of the parameter space of GC initial conditions. However, compared to earlier approaches, the integration is done on a star-by-star basis, which allows it to recover short-range physics such as stellar evolution and, more importantly for this paper, strong few-body interactions.

In \cmc, strong interactions are sampled from pairs of neighbouring stars and/or binaries according to their interaction rate. This interaction rate is computed from their relative cross section, defined as the area covered within the impact parameter that leads to a pericentre passage of $r_p$ if the binaries were taken to be point masses. For binary-single encounters, they fix $r_p=Xa$ (with $a$ the semimajor axis of the binary and $X$ a numerical constant), whereas for binary-binary encounters they use $r_p=X(a_0+a_1)$ (with $a_0$, $a_1$ the semimajor axes of the binaries). The authors find that using a prefactor of $X=2$ captures almost all the relevant energy-generating binary interactions \citep{FregeauRasio2007}. For this paper, we define `strong' interactions as those with $X\leq 2$. After sampling, these strong interactions are numerically resolved using \textsc{fewbody} \citep{fewbody}. Further details on the \cmc code can be found on \cite{cmc}.

We use the catalogue of 148 \cmc models of \cite{Kremer2020}, which is constructed by varying the number of stars $N$, viral radius $r_v$, metallicity $Z$, and galactocentric distance $R_G$. The choice of initial conditions represents a sampling of local-Universe GCs, and is $N\in [2, 4, 8, 16] \times 10^5$, $r_v \in [0.5, 1, 2, 4]\units{pc}$, $Z\in[0.0002, 0.002, 0.02]$, and $R_G\in[2, 8, 20]\units{kpc}$. This accounts for 144 models, plus four extra models with $N=3.2\times 10^6$, $R_G=20\units{kpc}$, $Z \in [0.0002, 0.02]$, and $r_v\in[1, 2]\units{pc}$.

\begin{figure}
    \centering
    \includegraphics[width=\columnwidth]{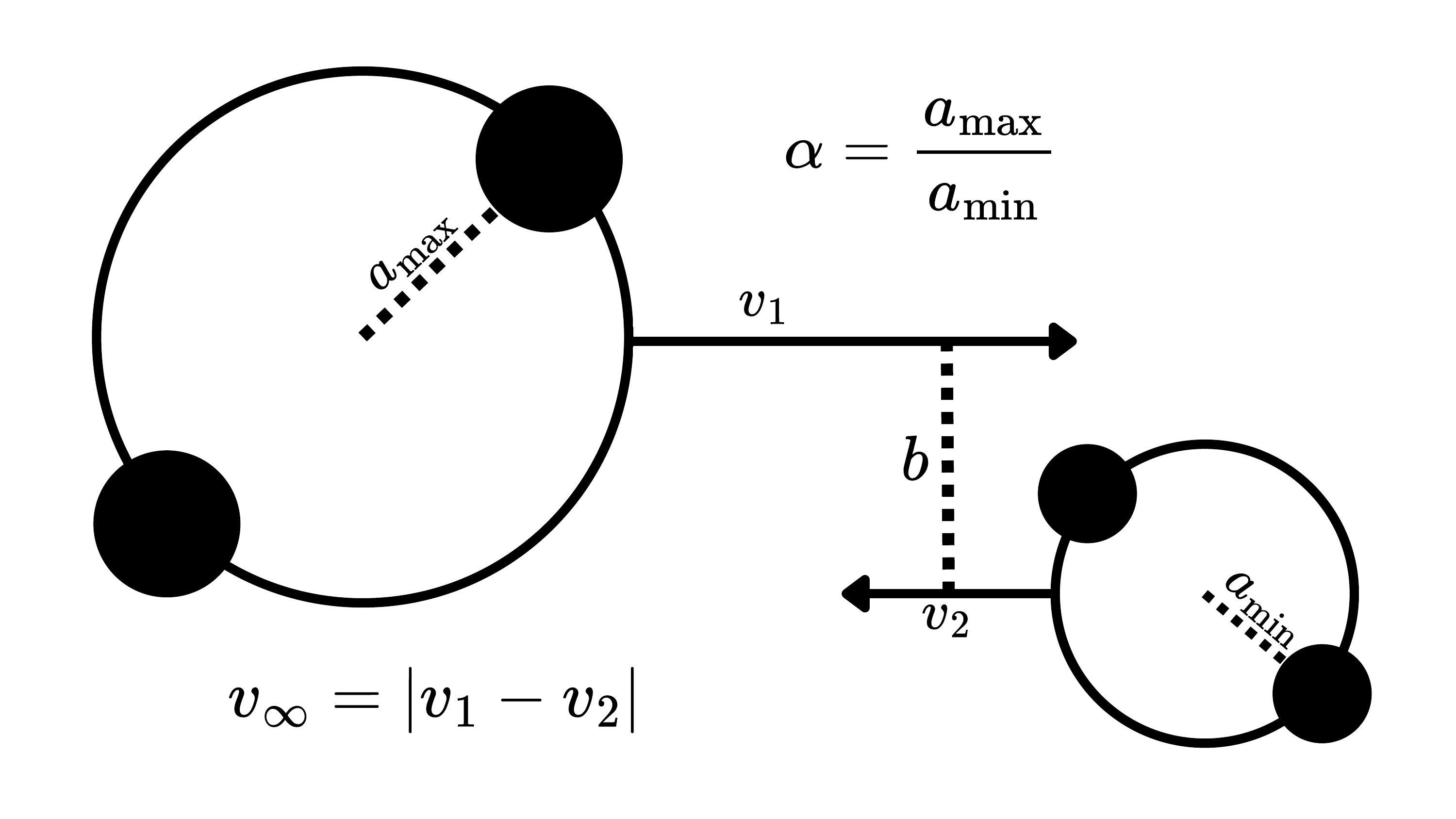}
    \caption{Schematic diagram of a BBH-BBH interaction with some of its relevant parameters.}
    \label{fig:binbin-diag}
\end{figure}

The advantage of using a Hénon-like approach to simulating GCs, as opposed to a direct \nbody integration, is that it allows us to perform a Monte Carlo sampling of the strong interactions. As described above, each strong encounter is considered as a scattering with fixed velocity, impact parameter, masses, semimajor axes, eccentricities and stellar type and radius. These parameters are shown schematically in Fig.~\ref{fig:binbin-diag}. The other 8 scattering parameters, which characterise the binaries' phases and orientations (for each binary, the inclination, argument of the periapsis, true anomaly and longitude of the ascending node), are randomly sampled. This allows us to re-simulate all the interactions with different parameters to increase our statistics without the need to perform costly simulations of entire GCs. We expand on this in the following section.

\subsection{Summary of \tsunami}
For each set of BBH-BBH interactions found in \cmc we ran ten different scattering experiments, sampling the phases and orientations of the binaries. This sampling was carried out using the \textsc{fewbody} code from \cite{fewbody}; however, we chose not to run the scattering simulations with \textsc{fewbody}, as it only includes the Newtonian and 2.5PN term. Instead, we used \tsunami \citep{Trani2019}, a state-of-the-art direct few-body integrator with post-Newtonian (PN) treatment of the equations of motion up to the 3.5PN term (inclusive), which is needed to accurately compute the dynamics before we identify a GW merger. \tsunami uses a logarithmic Hamiltonian and time-transformed leapfrog algorithm \citep{Mikkola1999a, Mikkola1999b} combined with a Bulirsh–Stoer extrapolation algorithm \citep{StoerBulirsch1980}. Each scattering was evolved in time until the interaction was resolved according to the criteria of Appendix~\ref{app:outcome}. Specifically, we assumed that a capture happens when the distance between two BHs is smaller than 10 times the sum of their Schwarzschild radii; we discuss this choice in Sect.~\ref{ssec:capcrit}. During an interaction, some configurations may lead to arbitrarily long-lived unstable intermediate states. In order to account for these cases, we label any scattering experiment that reached a simulation time of $t > 10^6 \units{AU^{3/2}} \,G^{-1/2} \units{\msun^{-1/2}} = 0.16\units{Myr}$ as unresolved. 

\section{BBH-BBH interactions}
\label{sec:bbhbbh}
In this section, we characterise the population of BBHs in star clusters and the mechanism that drives them to interact.

\subsection{Three-body BBHs}
The population of BBHs in a GC is not a simple fiducial evolution of the population of massive stellar binaries. After an initial settling phase, the cluster reaches mass segregation, where the heaviest components -- the BHs, after stellar evolution -- populate the cluster's core. Per \cite{Henon1961}, the energetic budget of a cluster is provided by the core, and thus the global properties of the cluster dictate the few-body BH dynamics in the central region. Specifically, the energy that fuels the cluster evolution is supplied by BBHs in the core, where a binary acts as a reservoir of negative binding energy, thus supplying (positive) energy to the rest of the cluster via interactions. According to \cite{Heggie1975,Goodman1984}, after core collapse there is a state of steady evolution where the energy demands of the cluster are balanced by the creation and hardening of binaries. When including BHs, it is the BBH formation in the BH subcluster that provides energy to the BH subcluster, which is in thermal contact with the stars \citep{BreenHeggie2013}. These binaries are formed dynamically from the encounter of at least three unbound BHs \citep{Heggie1975,Tanikawa2012, Atallah2024, Ginat2024}, so they are usually referred to as three-body binaries. When formed, these BBHs have large semimajor axes, $a$, with a distribution that formally diverges as $a\to\infty$ \citep{Heggie1975, Retterer1980}. This is because in a cluster, any two very separated BHs with sufficiently low relative velocity may be considered as bound. We avoid this issue by only considering `hard' BBHs, which have a (positive) binding energy $x=Gm_1m_2/(2a)$ that is greater than $m_\BH\sigma_{c, \BH}^2$, with $m_\BH$ and $\sigma_{c, \BH}$ the mean mass and velocity dispersion of BHs in the cluster core, respectively. The hardness of a binary can be quantified via its hardness ratio $z\equiv x/(m_\BH\sigma_{c, \BH}^2)$, so that hard binaries have $z \geq 1$. We only consider hard binaries because those that are less bound (i.e. `soft', $z < 1$) are easily dissociated by interactions with other unbound BHs. Opposite to this, \cite{Heggie1975} showed that hard BBHs tend to become more bound (harden) via these interactions, as a result imparting some kinetic energy to the third body and the centre of mass of the binary, which provides energy to the cluster. These BBHs end their life when their recoil kick after an interaction is enough to eject them from the cluster, when their semimajor axis shrinks enough such that they merge before the next interaction takes place, or when they undergo a GW capture during an interaction. In Section \ref{ssec:soft}, we discuss the effect of including soft BBHs interactions to the GW capture rate.

The population of BBHs in a GC can be divided in three separate categories: `primordial' BBHs\footnote{Not to be confused with BBHs of cosmological origin, whose existence has not yet been confirmed by observations.}, formed from the isolated evolution of a binary of massive stars without dynamical effects; `three-body' binaries, where the component BHs were originally unbound stars and the binary is assembled dynamically; and `exchange' binaries, that were formed dynamically but at least one of the component BHs was originally in a primordial binary.

It is reasonable to assume that a high primordial BBH fraction leads to a high BBH-BBH interaction rate. However, it is not explicitly shown in studies \citep{Banerjee2021b, Kremer2020, Zevin2019} that BBH-BBH interactions are indeed the result of primordial BBH. We first analyse the nature of the BBHs involved in BBH-BBH interactions. Specifically, this is done by checking whether the IDs of the BHs involved in the interaction were in binaries in the first snapshot of the simulation.

In our set of \cmc simulations, we find that less than 0.1\% of the strong\footnote{As explained in the previous section, we consider strong interactions as those with $r_p\leq 2(a_0+a_1)$} BBH-BBH interactions are among two primordial BBHs. Most of the interactions happen between two three-body binaries (84\%), with 15\% of interactions involving at least one exchange binary and $\sim$1\% involving two exchanged binaries. Therefore, we find that three-body BBHs dominate in BBH-BBH interactions. This can be explained by the continuous formation of BBHs near the hard-soft boundary and their immediate disruption in BBH-BBH interactions \citep{MarinPinaGieles2024}.

Therefore, in order to study the contribution of binary-binary interactions to the BBH merger rate in general and the eccentric GW production in particular, we focus on the dynamically assembled three-body BBHs. 

\subsection{BBH-BBH interaction rate}
The goal of this section is to understand the cluster processes that lead to close BBH-BBH interactions. Following \cite{MarinPinaGieles2024}, we assume that there is only one three-body BBH at any given moment in time, that continuously interacts with -- and ionises -- wide, recently formed three-body BBHs. This model predicts that most interactions among three-body BBHs will have large semimajor axis ratios, as these new three-body BBHs form closer to the hard-soft boundary, while the long-lived stable binary has undergone a process of hardening via binary-single interactions. 

\begin{figure}
    \centering
    \includegraphics[width=\columnwidth]{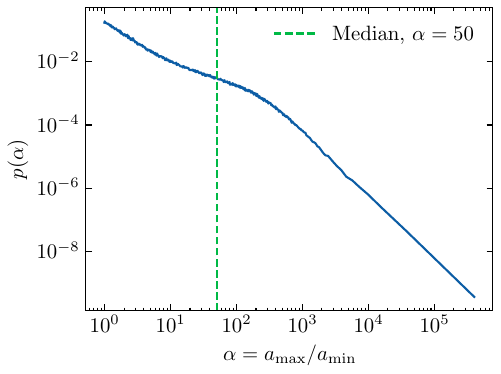}
    \caption{Probability density function of the ratio of the semimajor axes in BBH-BBH interactions, as calculated from the \cmc models.}
    \label{fig:interactions:SMAratio}
\end{figure}

This is indeed what we observe in the \cmc simulations, where we find a median semimajor axis ratio, $\alpha$, (defined as $\alpha=a_{max}/a_{min}$, with $a_{max}$ and $a_{min}$ the semimajor axis of the largest and smallest binary of the interaction, respectively) of $\alpha\sim50$. The complete distribution, shown in Fig.~\ref{fig:interactions:SMAratio}, can be explained by the interaction of a binary near the hard-soft boundary ($z\sim 1$, in the simulations we find a median $z\simeq 8$ for the BBH with $a=a_{max}$) with a stable, long-lived binary ($z\gg 1$, in the simulations we find a median $z\simeq 600$  for the BBH with $a=a_{min}$). We now assume that the distribution of energies of stable binaries follows some power-law $p(z)\propto z^{-b}$. From the models of \citet[][Fig.~14]{MarinPinaGieles2024}, the index of the exponent is $0.9\lesssim b \lesssim 1.4$. Then, assuming that $a_{max}$ is very large, the distribution of $\alpha$ must follow the same power-law dependence, $p(\alpha)\propto \alpha^{-b}$. This is indeed what is seen in Fig.~\ref{fig:interactions:SMAratio}, where the distribution can be described by a power law with exponent $b=1.1$.

\begin{figure}
    \centering
    \includegraphics[width=\columnwidth]{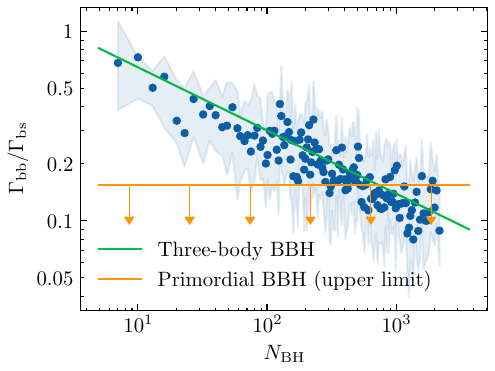}
    \caption{Ratio of the BBH-BBH interaction rate and the binary-single interaction rate as a function of the cluster's number of BHs at the time the interactions took place. The blue dots correspond to the median values obtained from the \cmc models (contour is the 64\% region). In green, the analytical prediction of Eq.~\ref{eq:scaling:ok}, which considers the interaction of three-body BBHs; yellow is the naïve prediction of Eq.~\ref{eq:scaling:wrong}, which only considers the interactions of primordial binaries.}
    \label{fig:interactions:BR}
\end{figure}

In order to quantify the relative importance of BBH-BBH interactions, we will compare the interaction rate of the stable BBH with other BBHs, $\Gamma_{bb}$, to the rate of interaction of this same central binary with single BHs, $\Gamma_{bs}$. As the BBH-BBH interaction rate is limited by dynamical BBH formation, it is roughly equal to the binary formation rate per unit volume and binding energy, $C$, which is \citep{GoodmanHut1993}
\eq{C(x) = \frac{2\pi^2}{3\sqrt{3}} \frac{n_\BH^3 G^5 m_\BH^4}{\sigma_{c, \BH}^{11}}\frac{z^{-9/2}}{(1+(5z)^{-1})(1+e^{-z})},}
with $n_\BH$ the central number density of the BHs. Note that this is different to the net formation rate $\bar{C}$ discussed in \citet[][see also \citealp{Atallah2024,Ginat2024}]{GoodmanHut1993}, which is the result obtained after considering binary ionisation due to binary-single interactions, as well as hardening and softening through the hard-soft boundary. We assume that these processes happen within the core volume, $V_c$, so the binary-binary interaction rate is
\eq{\label{eq:Gammabb} \Gamma_{bb}=m_\BH \sigma_{c, \BH}^2V_c\int_1^{\infty}C(z)\diff z}
For the binary-single interaction rate, we will use the $n$-$\Sigma$-$v$ approach \citep[as in e.g.][]{HillsDay1976} 
\eq{\label{eq:Gammabs} \Gamma_{bs}=n_\BH \avg{\Sigma_{bs} v}=n_\BH \int^\infty_0 \Sigma_{bs}(v)f_{Maxwell}(v)v\diff v,}
where $v$ is the relative velocity between single BH and BBH, and we assume equipartition between binaries and singles and a Maxwellian velocity distribution, $f_{Maxwell}(v)\propto v^2 \exp (- v^2/(3\sigma_{c, \BH}^2))$, where $\sigma_{c, \BH}$ is the (one-dimensional) central velocity dispersion of the BHs. We use the cross-section given by \cite[p.~217]{HeggieHut2003} 
\eq{\Sigma_{bs}=\frac{6 \pi G m_\BH a}{v^2}}
As discussed above, we assume that the stable binary has a binding energy  $\sim 50$ times larger than the hard-soft boundary. We can write the dependency on $\sigma_{c, \BH}$ in equations~\ref{eq:Gammabb} and \ref{eq:Gammabs} in terms of central density by using the definition of the core radius of the BH subsystem, $({4\pi G}/{9})\rho_0r_{c, \BH}^2=\sigma_{c, \BH}^2$ \citep[p.~71]{HeggieHut2003}. Finally, taking into account that for the isothermal and \cite{King1966} models with high concentration, the central density can be expressed in the average density within the core as $\rho_0 \simeq 1.9 \rho_{c, \BH}$, we can express the ratio of rates as
\eq{\frac{\Gamma_{bb}}{\Gamma_{bs}}=\frac{5.7}{N_{c, \BH}},}
where $N_{c, \BH}$ is the number of BHs in the core of the BH subcluster. For the two-mass model of \cite{BreenHeggie2013}, this quantity depends only on the number of BHs in the cluster \citep[equation~25]{MarinPinaGieles2024}, so that
\eq{\label{eq:scaling:ok}\frac{\Gamma_{bb}}{\Gamma_{bs}}\simeq 0.3 \, \lr{\frac{N_\BH}{10^2}}^{-1/3}}
This result implies that clusters with fewer BHs tend to have more binary-binary interactions with respect to binary-single interactions. In Fig.~\ref{fig:interactions:BR} we compare this analytic result to what we extract from the \cmc models and we find good agreement. We can compare this to the result that we would obtain if we were to assume that the interactions are among primordial binaries. If we consider a BH binary fraction of $f_b$, the ratio of rates would be \citep{MarinPinaGieles2024}
\eq{\label{eq:scaling:wrong}\frac{\Gamma_{bb}}{\Gamma_{bs}}=3.1f_b}
For a reference time of $\sim 100\units{Myr}$ after cluster formation, all our cluster simulations have $f_b$ below the initial binary fraction of $5\%$, likely due to BBH disruption at birth. In Fig.~\ref{fig:interactions:BR} we show the interaction rate ratio for the limiting case of $f_b=5\%$.
This prediction does not recover the $N_\BH$ dependence and thus cannot explain BBH-BBH interactions. We conclude that the high BBH-BBH interaction rate in the \cmc models can only be explained by three-body binaries. In the following sections, we study the outcomes of these interactions: GW captures and triple formation.

\label{ssec:scaling}

\section{GW captures in BBH-BBH interactions}
\label{sec:gwcaptures}
In this section, we study the GW captures in BBH-BBH interactions. We begin by deriving an analytical approximation to the probability of merging (Sect.~\ref{ssec:analytical}), which we compare to the set of our BBH-BBH scatterings sampled from realistic star cluster models (Sect.~\ref{ssec:singlescat}). In Sect.~\ref{ssec:weight}, we define the astrophysical weighting of our cluster models. In Sect.~\ref{ssec:mergerrate} and Sect.~\ref{ssec:ecc}, we obtain predictions for the merger rate and GW eccentricity distribution, as well as redshift dependence.

\subsection{Analytical approximation}
\label{ssec:analytical}
In order to do a full characterisation of the population of GW captures in BBH-BBH populations, it is necessary to first understand the mechanism that drives the captures in each interaction. In this section, we will present a simple analytical model for these captures. 

The gravitational four-body problem is chaotic, so in general we expect small differences in the initial conditions to grow exponentially. Therefore, our goal in this section is to provide distributions of outcomes for fixed initial parameters. 
We work in the gravitational focusing limit, i.e. $v_\infty \ll v_{crit}$, with $v_\infty$ the relative velocity at infinity and $v_{crit}$ the critical velocity for which the total energy in the centre-of-mass frame of the four-body system is zero. In the gravitational focusing limit, we can set the relative velocity at infinity to zero. This approximation is justified by the cluster simulations, in which only $\sim3$\% of the BBH-BBH interactions have $v_\infty/v_c>1$, and over half of them have $v_\infty/v_c<0.1$. In general, the masses of the BHs involved in BBH-BBH interactions are comparable, so we will work in the equal mass case in this Section. Then, the mass can be rescaled so the only parameters are $\alpha$ and $a_{max}$.

BBH-BBH scatterings can be separated in two different regimes. Scatterings with $\alpha \gtrsim 1$ are likely to be resonant with many intermediate states \citep{Zevin2019}, and thus have a high probability of resulting in a GW capture. During interactions with $\alpha \gg 1$, however, the tight binary may pass through the wider binary unaffected, or only have an encounter with one of its BHs. Therefore, interactions in this second regime are less likely to lead to GW captures of the wider binary, and the only possibility for GW mergers are those where $a_{min}$ is sufficiently small that the inner binary was already close to merger before the interaction. We expect the change in these two regimes to happen at some $\alpha_{crit}$. This can be measured from the scattering simulations by noticing that all mergers at $\alpha > \alpha_{crit}$ happen between the two BHs in the inner binary, whereas mergers at $\alpha < \alpha_{crit}$ may happen between any two BHs. This is explored in the following section.

In the case of resonant BBH-BBH scatterings, their intermediate states can be classified in two qualitatively different categories: `democratic states', where the distance between the four BHs is comparable, and `hierarchical states', where the distance between two of the BHs is much shorter than the other distances and the system behaves like an (unstable) hierarchical multiple. Democratic states are highly chaotic, so any information about previous orbital properties is erased. This allows us to model a resonant BBH-BBH interaction as a series of hierarchical states, each of which with a semimajor axis and eccentricity sampled from a certain distribution. This approach has been developed by \cite{Samsing2014} for the binary-single case. We begin by explaining its key arguments, and later extend the approach to the four-body case.

In the framework of \cite{Samsing2014}, each binary-single interaction has a mean number of hierarchical intermediate states (IMS) of $N_{IMS}=20$ \citep{Samsing2014, RandoForastier2024}. Each interaction is considered to end in a GW capture if at any given hierarchical intermediate state the eccentricity is such that, in a single pericentre passage, the energy radiated via GW emission is equal to the binding energy of the original binary. To leading order \citep{Hansen1972}, the critical eccentricity for the binary-single case, $e_{crit, 3}$, is
\eq{\label{eq:ecrit3} e_{crit, 3} \simeq 1-2.96\lr{\frac{Gm}{ac^2}}^{5/7}}
The probability of a capture in a single IMS is the fraction of the eccentricity distribution that is above $e_{crit}$. Assuming that the eccentricity of any intermediate state is drawn from a thermal distribution, we obtain that
\eq{p_{IMS} = \int_{\textit{e}_{crit, 3}}^1 2e\diff e = 1- e_{crit, 3}^2 \simeq2.96\lr{\frac{Gm}{ac^2}}^{5/7}.\label{eq:pIMS}}
The probability of merging in a binary-single interaction, $p_{merge, 3}$, is equal to the probability of a capture in any of the IMSs, which is 
\eq{\label{eq:pmerge3} p_{merge, 3} = 1-\lr{1-p_{IMS}}^{N_{IMS}} = 1-e_{crit, 3}^{2N_{IMS}}.} 

In the BBH-BBH case, we extend this framework by considering a critical eccentricity for a reference binary whose semimajor axis is equal to that of the smallest initial binary, $a_{min}$, with component masses equal to the mean mass of all four BHs, $\avg{m}$, i.e.
\eq{e_{crit} \simeq 1-2.96\lr{\frac{G\avg{m}}{a_{min}c^2}}^{5/7}}
We use $a_{min}$ as the reference semimajor axis because the tighter binary is more likely to merge. Furthermore, we assume that the merger probability has a correction factor $f(\alpha)$ with respect to the binary-single case, such that the probability of merging in a BBH-BBH interaction is
\eq{\label{eq:pmerge_pre}p_{merge} = f(\alpha) \lr{1-e_{crit}^{2N_{IMS}}}\simeq f(\alpha) \times 2N_{IMS}\lr{1-e_{crit}}}
This factor accounts for binary-binary interactions being more chaotic than binary-single interactions and thus having more complex resonances. Note that, because $e_{crit}\simeq 1$, increasing $f(\alpha)$ by a fixed factor is equivalent to increasing $N_{IMS}$. Thus, for comparison with the binary-single case, we fix $N_{IMS}=20$. Since $f(\alpha)$ is purely Newtonian, it can only depend on $\alpha$, and not the semimajor axes themselves. In the next section, we obtain a fit for $f(\alpha)$ from our BBH-BBH simulations. 

\subsection{Scattering simulations}
\label{ssec:singlescat}
We now compare the above analytical approximation for GW captures in BBH-BBH interactions against our detailed \tsunami simulations sampled from the \cmc cluster models. From the $\sim 2\times 10^6$ scatterings that we performed, $11\%$ were not resolved before the maximum integration time. From the resolved interactions, the most common outcome ($53\%$) is binary preservation\footnote{This depends on the cutoff value for $r_p$ in our cluster simulations, see Sect.~\ref{ssec:soft}}, where two BBHs remain after the scattering, followed by single ionisation ($26\%$) and triple formation ($21\%$). Mergers only account for $\sim 0.1\%$ of all resolved outcomes. For illustrative purposes, we show the animation of the trajectories of the BHs during two different interactions whose outcomes are GW mergers in here\footnote{\url{https://youtu.be/HHl85a92v0M}} (for $\alpha\simeq 1$) and here\footnote{\url{https://youtu.be/4Wu-E_BFPbQ}} (for $\alpha> 1$).

\begin{figure}
    \centering
    \includegraphics[width=\columnwidth]{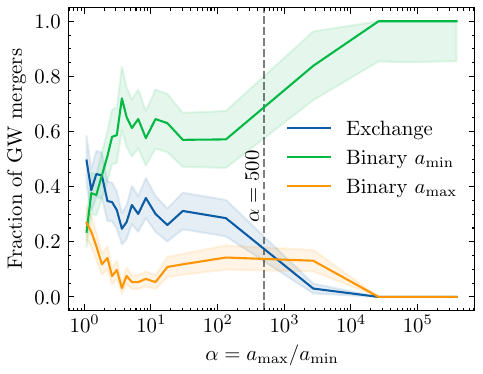}
    \caption{Fractions of GW mergers in BBH-BBH interactions as a function of the initial $\alpha$. In green, mergers between the BHs of the smaller BBH; in yellow, mergers between the BHs of the larger BBH; in blue, mergers of any other two BHs. Each point represents the mean among multiple runs binned on their value of $\alpha$, the shaded areas correspond to the standard deviation of the mean.}
    \label{fig:singlescat:outcome-alpha}
\end{figure}

\begin{figure}
    \centering
    \includegraphics[width=\columnwidth]{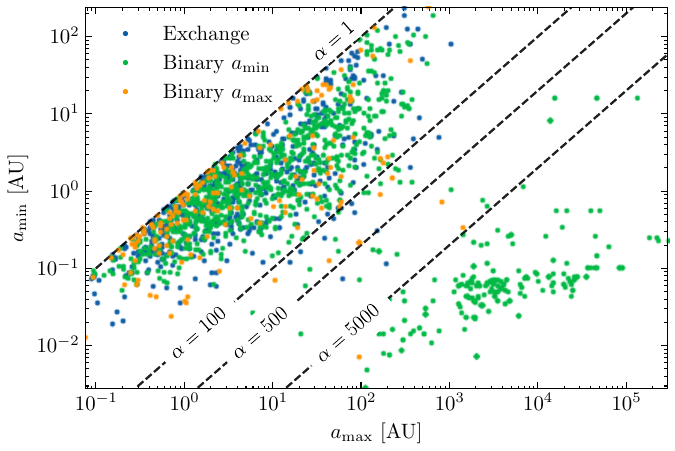}
    \caption{GW mergers in BBH-BBH interactions in the initial $a_{max}$, $a_{min}$ parameter space. In green, mergers between the BHs of the smaller BBH; in yellow, mergers between the BHs of the larger BBH; in blue, mergers of any other two BHs.}
    \label{fig:singlescat:outcome-a}
\end{figure}

As predicted in the previous section, we find two different qualitative behaviours in GW mergers depending on their $\alpha$ (see Fig.~\ref{fig:singlescat:outcome-alpha} and Fig.~\ref{fig:singlescat:outcome-a}). Interactions with $\alpha < \alpha_{crit}$ can have complicated resonances and captures can happen between any two BHs, although we find a bias towards the merging of the smaller binary. Combinatorics would predict that, in fully mixed resonant encounters of four BHs, mergers of the smaller binary represent only $1/6$ of all captures, with the larger binary representing another $1/6$ and the rest, $2/3$. However, this would only be the case if the systems are strongly resonant with equal masses and SMA. All our models have $\alpha>1$, which leads to a bias of the smaller binary merging, although we do see the expected ratios at $\alpha \simeq 1$. Also, about 30-50\% of the interactions are non-resonant, meaning that they merge after only one IMS \citep{Trani2024b, RandoForastier2024}. This mechanism biases the outcome ratios to overrepresent mergers with the input configuration.

Furthermore, we find that the ratio of merger types is roughly constant for $3 \lesssim \alpha \lesssim 100$, although the uncertainties allow for a slight increase or decrease. This may be partly explained by a bias towards the merging of the smaller binary due to its shorter merger timescale. A binary in isolation with a SMA of $a$ and a critical eccentricity $e_{crit, 3}$ (equation~\ref{eq:ecrit3}) has a GW inspiral timescale \citep{Peters1964}
\eq{t_{GW}(a)\sim\frac{a^4}{\beta}\lr{1-e^2_{crit, 3}(a)}^{7/2}\propto a^{1.5},}
where $\beta$ is a constant that only depends on the masses of the system. Thus, the ratio of the merger timescales of the initial binaries when they are at the critical eccentricity is
\eq{\frac{t_{GW}(a_{max})}{t_{GW}(a_{min})}\propto \alpha^{1.5}}
In comparison, the duration of the interaction $t_{res}$ can be estimated as $N_{IMS}$ times the duration of the hierarchical state, each with an orbital period corresponding to a binary of $a\sim a_{max}$, i.e.
\eq{t_{res}\sim N_{IMS} \sqrt{\frac{a_{max}^3}{G m}}}
For highly eccentric binaries, $t_{GW}(a_{min}) \ll t_{res}$, but $t_{GW}(a_{max}) \gtrsim t_{res}$. In this case, the tight binary can merge in between intermediate states, while the larger binary may be perturbed by the other BHs before it can inspiral. Thus, these different timescales introduce a bias towards the merging of the smaller binary.

On the other side, interactions with $\alpha > \alpha_{crit}$ generally only undergo mergers of the smaller binary. In this limit, the interaction is not truly resonant, so the only possible mergers are due to the smaller binary being so tight that it is driven to coalesce either due to GW emission alone or via a weak perturbation of the eccentricity in a direct interaction\footnote{See \url{https://youtu.be/4Wu-E_BFPbQ} for an example animation of this case} \citep{HeggieRasio1996}. We find a transition at $\alpha_{crit}\sim 500$; combined with the steep $\alpha$ distribution, most mergers ($89\%$) happen at $\alpha < \alpha_{crit}$.

\begin{figure}
    \centering
    \includegraphics[width=\columnwidth]{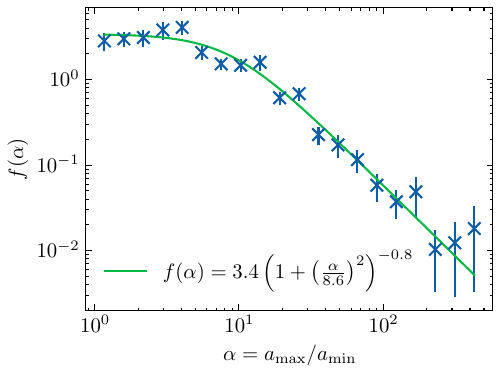}
    \caption{Correction factor $f(\alpha)$ (defined in Sect.~\ref{ssec:analytical}) as a function of the semimajor axes ratio $\alpha$. A value of $f(\alpha)>1$ implies that the binary-binary interaction is more likely to trigger a GW merger than a binary-single interaction where the binary has semimajor axis $a_{min}$.}
    \label{fig:singlescat:f-alpha}
\end{figure}

For encounters with $\alpha<\alpha_{crit}$, we compute $f(\alpha)$ by substituting the merger probability, obtained in the \tsunami simulations, in equation \ref{eq:pmerge_pre}. We show the result in Fig.~\ref{fig:singlescat:f-alpha}, where we plot $f(\alpha)$ (and corresponding uncertainties) in logarithmic bins, which we fit with the following function
\eq{f(\alpha)=\lambda_1\lr{1+\lr{\frac{\alpha}{\lambda_2}}^2}^{-\lambda_3/2}.}
This function has the desired properties of a flattening at $\alpha\to 1$ and a power-law scaling at large $\alpha$. Specifically, the value of $\lambda_1$ fixes the vertical scaling, $\lambda_2$ marks the transition from flattening to power-law, and $\lambda_3$ fixes the power-law index at large $\alpha$ ($f(\alpha)\propto \alpha^{-\lambda_3}$). We perform a non-linear least-squares fit, for which we obtain $\lambda_1=3.4$, $\lambda_2=8.6$, and $\lambda_3=1.7$.

This shows that binary-binary interactions are more likely to trigger a GW capture when their semimajor axes are comparable, and decrease with $\alpha$. In the case of $\alpha=1$, we recover the result of \cite{Zevin2019} where binary-binary interactions are about $\sim 5$ times more likely to trigger a GW capture than binary-single interactions for any value of the semimajor axis. However, for $\alpha > 16$ (which is about $65\%$ of our interactions), the capture probability is smaller than for an equivalent binary-single interaction. As a summary, the probability of merging in a BBH-BBH interaction with $\alpha < 500$ is
\eq{\label{eq:pmerge} p_{merge} \simeq 0.034\lr{\frac{\avg{m}}{20\units{\msun}}}^{5/7}\lr{\frac{a_{min}}{0.1\units{AU}}}^{-5/7} \lr{1 + \lr{\frac{\alpha}{8.6}}^2}^{-0.83}}
In the cluster simulations, BBH-BH interactions are about $\sim 7$ times more frequent than BBH-BBH. However, equations \ref{eq:pmerge3} and \ref{eq:pmerge} allow us to compute the probability of capture for all the BBH-BH and BBH-BBH scatterings in the cluster simulations, respectively. We find that each BBH-BBH interaction is about $\sim 3$ times more likely to trigger a merger than a BBH-BH interaction, so the total rate of captures in BBH-BBH interactions is about half that in BBH-BH interactions. This is due to both the contribution of the $f(\alpha)$ term and the different distribution of $a_{min}$ in BBH-BBH interactions with respect to $a$ in BBH-BH interactions.

\subsection{GC population model weighting}
\label{ssec:weight}
The grid of cluster models presented above sufficiently samples the parameter space of current-day Milky Way GCs \citep{Kremer2020}; however, in order to extrapolate our predictions to the GC population in the Universe, we must weight each model according to how common the distribution of its initial parameters is (e.g. in general, there are fewer clusters with high masses than clusters with lower masses). We follow a similar approach as in previous studies on the topic \cite[e.g.][]{Kremer2020, Antonini2023, YeFishbach2024}. Firstly, we assume that the distribution of clusters in the Universe follows the quasi-separable distribution 
\eq{\phi(M_0, r_v, R_G, Z, z) \equiv \phi_{\textit{M}_0}(M_0) \phi_{\textit{r}_v}(r_v) \phi_{\textit{R}_G}(R_G) \phi_{\textit{Z}}(Z, z)\phi_{\textit{z}}(z),}
where each of the $\phi_{\it i}$ is a normalised probability density function. From these, we assume a Schechter-like initial cluster mass function \citep{Schechter1976}
\eq{\label{eq:phiM0}\phi_{\textit{M}_0}(M_0)\propto M_0^{-2}e^{-M_0/M_S},}
with minimum mass $M_{min}=10^4\units{\msun}$, maximum mass $M_{max}=2\times10^7\units{\msun}$, and cutoff mass $M_S=10^{6.2}\units{\msun}$ \citep{AntoniniGieles2020a}. We can use this to compute the number density of GCs formed across cosmic time, $n_{GC, 0}$. Assuming that the cluster mass formed per unit volume is $\rho_{GC, 0}\simeq 2.4 \times 10^{16}\units{\msun}\units{Gpc^{-3}}$ \citep{Antonini2023}, we obtain
\eq{n_{GC, 0}=\frac{\rho_{GC, 0}}{\int^{M_{max}}_{\textit{M}_{min}} M_0\phi_{\textit{M}_0}(M_0)\diff M_0} \simeq 5\times 10^{11}\units{Gpc^{-3}}.}
We further weight our models with a log-normal distribution for the metallicity
\eq{\phi_Z(Z, z)\propto \frac{1}{Z}\exp \lr{\frac{- \log^2_{10} \lr{Z/\avg{Z(z)}}}{2\sigma_Z^2}} \quad ; \quad Z \in [0, 1],}
with parameters $\sigma_Z=0.25$ and $\avg{Z(z)}$ as given by \citep{MadauFragos2017}. 
\eq{\log_{10} \avg{Z(z)/Z_\sun} \simeq 0.153 -0.074z^{1.34}.}
We use a normal distribution for the cluster virial radii as in \cite{FishbachFragione2023}
\eq{\phi_{\textit{r}_v}(r_v) \propto \exp \lr{-\frac{(r_v-\mu_{\textit{r}_v})^2}{2\sigma_{\textit{r}_v}^2}} \quad ; \quad r_v \in [0.5, 4],}
with mean $\mu_{\textit{r}_v}=2\units{pc}$ and standard deviation $\sigma_{\textit{r}_v}=2\units{pc}$. For the cluster formation rate, we follow the procedure in \cite{Antonini2023} and sample the formation redshift of our cluster models, $\phi_{\it z}(z)$, from the fiducial semi-analytical galaxy formation model of \cite{ElBadry2019}, shown in Fig.~\ref{fig:result:z}.
Finally, we weigh all three Galactocentric radii equally.

\subsection{Merger rate}
\label{ssec:mergerrate}
\begin{figure}
    \centering
    \includegraphics[width=\columnwidth]{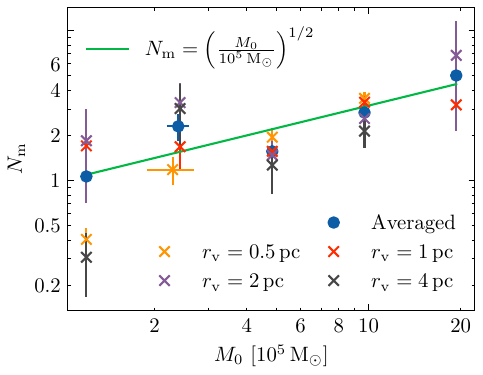}
    \caption{Number of mergers per cluster due to GW captures in BBH-BBH interactions as a function of initial cluster mass, separated by initial $r_v$ (crosses) and averaged over initial $r_v$ (points).}
    \label{fig:result:M0}
\end{figure}

In order to obtain the merger rate, we have to account for initial cluster masses below and above the mass range in our catalogue. We do so by following a procedure similar to \cite{Kremer2020}. The number of GW captures in BBH-BBH interactions $N_m$, shown in Fig.~\ref{fig:result:M0}, can be described by a power-law dependence on the initial cluster mass. Within the uncertainties, this is independent on the cluster's initial concentration (also shown in Fig.~\ref{fig:result:M0}) and metallicity. The data can be well described by
\eq{\label{eq:Nm} N_m = \lr{\frac{M_0}{10^5\units{\msun}}}^{1/2}.}
We note that this scales slower than linearly and, therefore, the merger efficiency per unit cluster mass is higher in low-mass clusters.  This is a direct consequence of the argument of Sect.~\ref{ssec:scaling}, where we obtained that BBH-BBH interactions were more common in clusters with fewer BHs. If we average $N_m$ with equation~\ref{eq:phiM0}, we obtain a mean number of mergers per cluster across the whole initial mass function of $\avg{N_m}\simeq 0.6$. It is important to note, however, that this result depends on the adopted $M_{min}$ (and to a smaller degree, on $M_{max}$).

We use this result to compute the comoving merger rate $\mathcal{R}(z)$ by imposing that its integral over time is equal to the (mass-weighted) number of mergers per cluster, $\avg{N_m}$, times the number density of GCs formed across cosmic time, $n_{GC, 0}$, such that
\eq{\label{eq:ratenorm} \avg{N_m} n_{GC, 0}=\int^{t_0}_0 \mathcal{R}(z(t)) \diff t,}
where $z(t)$ is the function that converts from lookback time to redshift and $t_0$ is the age of the Universe, both assuming the \cite{Planck2018} cosmology. 

\begin{figure}
    \centering
    \includegraphics[width=\columnwidth]{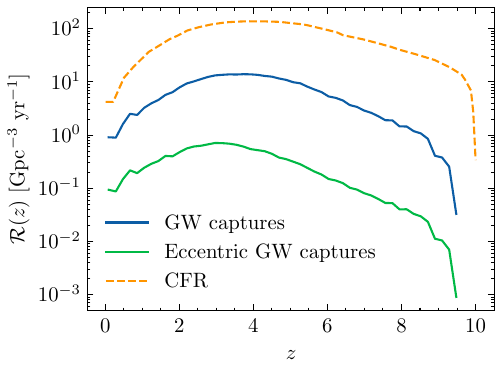}
    \caption{In blue, merger rate due to captures in BBH-BBH interactions as a function of redshift. In green, the same rate but only considering eccentric captures ($e>0.1$, measured at $f_{22}=10\units{Hz}$ as explained in Sect.~\ref{ssec:ecc}). In yellow, the cluster formation rate of \cite{ElBadry2019}, in the same units as the merger rate.}
    \label{fig:result:z}
\end{figure}

The merger rate as a function of redshift for GW captures in BBH-BBH interactions is shown in Fig.~\ref{fig:result:z}. These interactions usually happen early in the cluster evolution when the cluster is at its densest. Furthermore, the time from BBH formation to merger is short for GW captures (compared to ejected mergers), so their distribution with redshift is a close tracer of the cluster formation history. This prediction should be independent of the specific details of cluster evolution, so the detection of GWs with non-zero eccentricity can be used to constrain cluster formation models \citep{RomeroShaw2021a}. In our case, we obtain that the peak rate of GW captures in BBH-BBH interactions happens at $z\simeq 3.7$. This result can be compared to the redshift distribution of the isolated channel, where the merger rate is predicted to follow the star formation rate \citep[SFR,][]{MadauDickinson2014} with some delay. At $z\simeq 0$, the slope of our assumed cluster formation rate \citep[CFR,][]{ElBadry2019} is similar to that of the SFR, so that the slope of the merger rate as a function of redshift cannot be used to separate the dynamical and the isolated formation channels. However, we predict a peak rate at a higher redshift than the SFR, whereas the isolated channel predicts a peak at $z\sim 2$ \citep{Santoliquido2020}. This can be used to separate both channels, although it is a general prediction of cluster dynamics and not specifically of BBH-BBH interactions. 

For GW captures in BBH-BBH interactions, we predict a local merger rate of
\eq{\mathcal{R}(z\simeq 0) = 0.9\units{Gpc^{-3}}\units{yr^{-1}}}
Assuming a total rate of BBH mergers of $23.9 \units{Gpc^{-3}}\units{yr^{-1}}$\citep{LVK2023}, we estimate that the BBH-BBH channel contributes to about $\sim 4\%$. 

\subsection{Eccentricity distribution}
\label{ssec:ecc}
To identify the GWs produced in a BBH-BBH interaction, it will be sufficient to study the dynamics of the two merging BHs as if they were isolated, because their relative separation is orders of magnitude smaller than the separation to the other two BHs in the interaction \citep[but see][for the possibility of detecting phase shifts in the GW signal due to a nearby companion]{2024arXiv240305625S}. Even with this simplification, the inclusion of PN terms causes the orbits to deviate from a Keplerian parametrisation, and therefore the eccentricity is ill-defined. There exist multiple ways of defining eccentricity in General Relativity in such a way that it converges to the Newtonian value in the weak-field regime \citep{Shaikh2023}. In PN theory, we can use a quasi-Keplerian parametrisation with three different values of the eccentricity: temporal $e_t$, radial $e_r$, and angular $e_\phi$; which are only equal in the Newtonian regime. We discuss the nomenclature for these eccentricities in Appendix~\ref{app:ecc}. In this section we report on $e_t$; by convention, we will simply call it $e$, although keeping the above discussion in mind.

A further complication of this quasi-Keplerian parametrisation is the inclusion of non-conservative PN terms, which start at 2.5PN. These terms describe the decay and circularisation of the orbit via GW emission, which implies that the eccentricity is no longer a constant of motion. The production of GWs is very efficient at circularising the binary, so it becomes necessary to define a point in the orbit at which to report the eccentricity. To overcome this issue, we report the value of the eccentricity when the frequency of the 22-mode, $f_{22}$, is equal to 10\,Hz. This is usually taken to be the moment at which a quasi-circular binary enters the LIGO-Virgo-Kagra frequency band. The choice of reference frequency ($f_{22}$) is motivated by recommendations in the literature that aim to standardise the definition of eccentricity \citep{RamosBuades2022, Shaikh2023, Vijaykumar2024}. Our definition matches that of \cite{Shaikh2023}, used in GW data analysis \citep{Vijaykumar2024}. However, previous work in the field uses the peak GW frequency \citep{Wen2003} as their reference value. Both these definitions agree only at small $e$ \citep{Vijaykumar2024}. We compute $f_{22}$ as
\eq{f_{22}\equiv 2f_{orb}=\frac{\Phi n}{\pi}}
where $f_{orb}$, $\Phi$ and $n$ are the orbital frequency, angle of return to the periapsis and mean motion, respectively. The last two quantities are computed in 3PN approximation as \citet[their equations~25c,k]{Memmesheimer2004}. 

\begin{figure}
    \centering
    \includegraphics[width=\columnwidth]{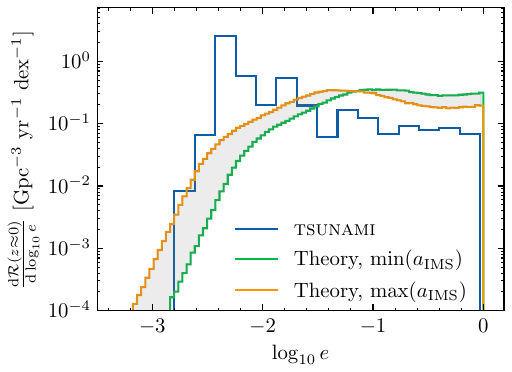}
    \caption{Merger rate due to captures in BBH-BBH interactions as a function of the eccentricity (measured at $f_{22}=10\units{Hz}$, as explained in Sect.~\ref{ssec:ecc}). In blue, the results obtained from the \tsunami simulations. In green and yellow, the analytical model of Sections~\ref{ssec:analytical} and \ref{ssec:ecc}, obtained from the \cmc data, and computed from the minimum and maximum $a_{IMS}$ (equation~\ref{eq:aIMS}), respectively. For clarity, the area between these lines is coloured in light grey.}
    \label{fig:result:ecc}
\end{figure}

In order to obtain the eccentricity distribution of captures in BBH-BBH interactions as observable by an Earth-based GW detector, we use the list of mergers found in the above scattering experiments. We associate a weight with each merger, as described in Sect.~\ref{ssec:weight}, to take into account the distribution of cluster parameters across the Universe, with the total rate normalised as in equation~\ref{eq:ratenorm}. The results are shown in Fig.~\ref{fig:result:ecc}.

For comparison, we also plot the simple analytical model of Sect.~\ref{ssec:analytical}. During a four-body hierarchical intermediate state, the inner binary has a semimajor axis of $a_{IMS}$. If we model this IMS as a point-like inner binary with two orbiting BHs, due to conservation of energy we obtain a range for $a_{IMS}$ of
\eq{\label{eq:aIMS} a_{IMS}\in [1, 5] \lr{\frac{1}{a_{min}}+\frac{1}{a_{max}}}^{-1}.}
We obtain the predicted eccentricity distribution by sampling a thermal distribution of eccentricities above $e_{crit}$ for each BBH-BBH scattering. Then, we evolve these eccentricities (starting at $a=a_{IMS}$) using 2.5PN orbit-averaged equations \citep{Peters1964} until the GW emission reaches $f_{22}=10\units{Hz}$. In Fig.~\ref{fig:result:ecc} we show these results, weighted both according to their $p_{merge}$ (equation~\ref{eq:pmerge}) and their parameter distribution (Sect.~\ref{ssec:weight}).

As can be seen, the distribution of eccentricities is roughly flat in $\log_{10} e$ between $-2.5$ and $0$, with $\sim10\%$ of mergers having $e>0.1$; and the simple model can roughly reproduce the simulations. Our results for the eccentricity distribution can be used directly as a prior for Numerical Relativity studies in eccentric waveforms. 

The above results assume that $z \simeq 0$, so that $f_{22}$ is the same in the source frame and the detector frame. If we fix the eccentricity to $f_{22}=10\units{Hz}$ in the detector frame, we are actually measuring the eccentricity at higher frequencies in the source frame. The effect of this is to shift the eccentricity distribution towards lower values of $e$ and reduce the fraction of eccentric mergers for higher $z$. In Fig.~\ref{fig:result:z}, we show the rate of mergers with $e>0.1$ at $f_{22}=10\units{Hz}$ in the detector frame, as a function of redshift.

Note that, since our definition of eccentricity is different from the commonly used \cite{Wen2003}, there is not a clear comparison between our results and e.g. \cite{AntoniniGieles2020a}. However, both definitions agree at low $e$ \citep{Vijaykumar2024}, so that the fraction of mergers with $e > 0.1$ is roughly independent of the eccentricity definition. We find that $\sim 10\%$ of our mergers have $e>0.1$, whereas \cite{AntoniniGieles2020a} find that about half of their GW captures in binary-single scatterings have  $e>0.1$. A similar fraction of eccentric mergers in binary-single scatterings was found in 
\citet{Trani2024b}. This discrepancy can be attributed to both the different methods to integrate the orbit as well as the intrinsic differences between binary-single and binary-binary scatterings.

\section{Triple formation and von~Zeipel-Lidov-Kozai oscillations}
\label{sec:triples}
In the Newtonian point-mass approximation, the possibility that a binary-single interaction leads to the formation of a stable triple is zero \citep[p. 211]{HeggieHut2003}. This can be seen heuristically by considering time reversibility: an isolated three-body system that is stable in the future needs to be stable in the past due to the symmetry of the Newtonian equations of motion under time reversal. Therefore, binary-binary interactions are the dominant dynamical process for stable triple formation. In this section, we explore dynamically assembled triples using our scattering simulations. As explained in Table~\ref{tab:outcome}, we classify as triples any endstate that verifies the \cite{MardlingAarseth2001} stability criterion \citep[but see][for more precise dynamical stability criteria]{2022ApJ...938...18L,h22022ApJ...939...81H,h12023ApJ...943...58H}.

\begin{figure}
    \centering
    
    \subfloat[Semimajor axes]{%
      \includegraphics[width=\columnwidth]{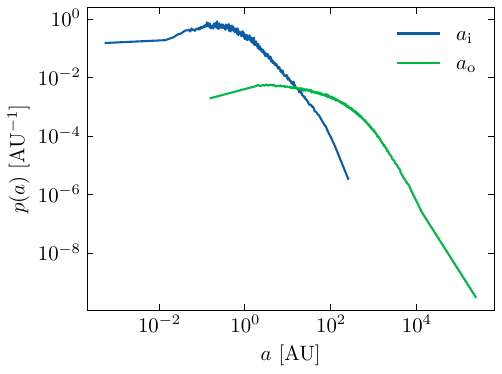}%
    }
    
    \subfloat[Eccentricity]{%
      \includegraphics[width=\columnwidth]{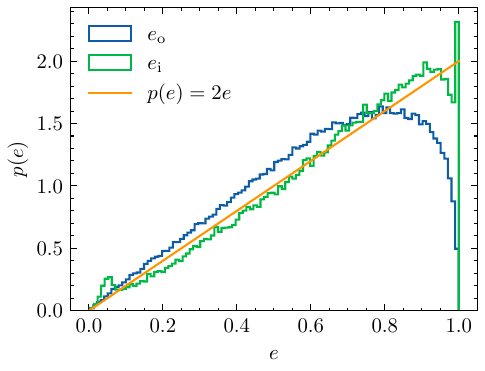}%
    }
    
    \caption{Probability density function of the semimajor axes (top) and eccentricity (bottom) for the inner and outer binaries of BH triples dynamically assembled via BBH-BBH interactions.}
    \label{fig:triple:sma}
\end{figure}

In Fig.~\ref{fig:triple:sma}, we show the semimajor axes of the inner ($a_i$) and outer ($a_o$) binaries of the dynamically assembled BH triples in our models. Generally, these triples have large $a_o$, with a median semimajor axis ratio between outer and inner binary of $\sim120$, and a mode of $\sim14$. The distribution of eccentricities, also shown in Fig.~\ref{fig:triple:sma}, is close to thermal for both the inner ($e_i$) and outer ($e_o$) binaries. However, there is a lack of high-eccentricity outer binaries; this is because such triples are unstable \citep{MardlingAarseth2001}. A similar distribution for dynamically formed triples in low-mass star clusters was found in \citet{Trani2022}.

Since dynamically assembled triples generally have large $a_o$, they are easily destroyed or destabilised in triple-single interactions. We can estimate the timescale as $t_{ts} = 1/(n \Sigma_{ts} v)$. Here, $n$ is the number density of stars around the triple, that we assume equal to the mean number density in the cluster core; $v$ their relative velocity, which we assume equal to the root mean square density-weighted velocity at the cluster centre; and $\Sigma_{ts}$ the cross-section of resonant triple-single interactions, as given by \cite[][below equation~26]{AntogniniThompson2016} 
\eq{\Sigma_{ts} = \pi a_o^2 \lr{\frac{\hat{a}}{\hat{v}^{3/2}}+\lr{\frac{\hat{v}^2}{\hat{b}}+\frac{\hat{v}^6}{\hat{c}}}^{-1}}}
with parameters\footnote{Note that there is a typo in Fig.~9 in \cite{AntogniniThompson2016}, when they quote their best fitting parameters as [sic] $\hat{\sigma}_a \simeq 0.257$, $\hat{\sigma}_b \simeq 0.044$, $\hat{\sigma}_c \simeq 8.89\times 10^{-3}$; these should be $\hat{\sigma}_b \simeq 0.257$, $\hat{\sigma}_c \simeq 0.044$, $\hat{\sigma}_a \simeq 8.89\times 10^{-3}$.} $\hat{a}=8.89\times 10^{-3}$, $\hat{b}=0.275$, and $\hat{c}=0.044$, and the velocity written as $\hat{v}=v/v_{crit}$.

After a triple is dynamically assembled, the inner binary may change its orbital inclination and eccentricity due to the interaction with the outer component, an effect that is known as von~Zeipel-Lidov-Kozai (ZLK) oscillations \citep{zeipel1910,Lidov1962, Kozai1962}. If the eccentricity reaches a sufficiently high value, the inner binary may be driven to merge due to GW emission at pericentre. We will now estimate the relative contribution of this channel to the observable GW rate.

We begin by assuming that the ZLK effect is only relevant for triples with a sufficiently large inclination of the outer to the inner orbit, $i_0$. For triples with $\abs{\cos i_0} < 3/5$, the eccentricity of the inner binary can reach a value of $e_{max}^\text{DA}$, given by \cite{Naoz2016, Mangipudi2022}
\eq{e_{max}^\text{DA} = \sqrt{1-\frac{5}{3}\cos^2i_0}.}
The timescale for this process, $t_{ZLK}$, is \citep{Liu2023}
\eq{t_{ZLK}=2\pi \sqrt{\frac{a_i^3}{G\lr{m_{i, 1}+m_{i, 2}}}}\frac{m_{i, 1}+m_{i, 2}}{m_o}\lr{\frac{a_o}{a_i}}^3\lr{1-e_o^2}^{3/2},}
where $m_{i, 1}$ and $m_{i, 2}$ are the masses of the components of the inner BBH and $m_o$ is the mass of the outer BH. On average, a triple undergoes a number of ZLK cycles, $N_{ZLK}$, before being disrupted, with $N_{ZLK}=t_{ts}/t_{ZLK}$. We will assume that a merger occurs if the maximum eccentricity reaches a critical value $e_{crit}$, which we define similarly to the previous section
\eq{e_{crit} \simeq 1-2.96\lr{\frac{G^5 m_{i,1}^2m_{i,2}^2(m_{i,1}+m_{i,2})}{2a_i^5c^{10}}}^{1/7}}
Therefore, the inner binary of a triple merges due to the ZLK mechanism if $\abs{\cos i_0} < 3/5$, $N_{LK}\geq 1$ and $e_{max}^\text{DA} \geq e_{crit}$. We find that, on average a fraction  $0.14\%$ of the dynamically assembled triples in our simulations will have time to merge due to ZLK interactions before they are destabilised by an interaction. Even though there are three orders of magnitude more dynamically assembled triples than GW captures ($21\%$ and $0.1\%$ of the resolved BBH-BBH outcomes in our simulations before model weighting, respectively), the GW capture rate is $\sim3$ times higher than the ZLK induced merger rate in dynamical triples. Then, taking the merger rate of Sect.~\ref{ssec:mergerrate}, we obtain that for dynamically assembled triples, $\mathcal{R}=0.3\units{Gpc^{-3}}\units{yr^{-1}}$. This is compatible with previous studies, e.g. \cite{Martinez2020}.

We note that this result only applies to triples that have formed in BBH-BBH scatterings, are stable, and then merged due to their secular ZLK evolution. Previous work using $N$-body models \citep[e.g.][]{Banerjee2018b} agrees with our result that this channel is subdominant, instead arguing for mergers in dynamically formed unstable triples. This is equivalent to what we define as GW captures during BBH-BBH and BBH-BH encounters, as the interaction is not formally resolved if the triple is unstable. 

\section{Discussion}
\label{sec:discussion}
\subsection{Capture criteria}
\label{ssec:capcrit}
In this paper, we have assumed that a capture is inevitable when the distance between two BHs is smaller than 10 times the sum of their Schwarzschild radii. A trade-off exists; if we considered a larger cutoff value, we could mislabel hyperbolic encounters as mergers. For a smaller cutoff, we would be integrating the PN equations of motion beyond their validity range. For quasi-circular orbits this is mostly fine, as the collision time between two stellar-mass BHs in such an orbit is of the order of seconds. However, when introducing highly eccentric or hyperbolic orbits, the merging time can be arbitrarily long. Some authors have used the orbital semimajor axis as the cutoff criteria, e.g. in the case of \cite{Zevin2019}, the authors classify a merger if the semimajor axis is smaller than 10 times the sum of the BHs' Schwarzschild radii, but as explained above, that allows the periapsis to be arbitrarily small, breaking the PN expansion.
Furthermore, there can be hyperbolic encounters of two BHs \citep{Capozziello:2008ra, DeVittori:2012da, Bini:2021jmj, Grobner:2020fnb, Garcia-Bellido:2017knh} such that their orbit occurs in the strong field regime but their endstate is not a merger \citep{Hopper:2022rwo, Damour:2014afa, Jaraba:2021ces}\footnote{There have been recent searches for hyperbolic encounters in \cite{Bini:2023gaj, Morras:2021atg}, although no such events have been found in LVK data to date.} The definition of a capture criteria is, then, beyond the scope of the PN expansion, and one must make use of full Numerical Relativity. This is computationally expensive and well beyond the scope of this work, although limited simulations in Numerical Relativity suggest that the energy change at our cutoff distance is such that, for the GW captures in this work, the merger cannot be avoided.

\subsection{Soft binaries and wide scatterings}
\label{ssec:soft}

In a cluster, the definition of what counts as strong scattering is somewhat arbitrary. As explained in Sect.~\ref{ssec:cmc}, we consider all interactions that have $r_p=X(a_0+a_1)$, with $X$ an arbitrary value that we fix at $X=2$. Increasing $X$ would include more scatterings whose most likely outcome would be binary preservation. As such, the outcome ratios in Sect.~\ref{ssec:singlescat} are dependent on the value of $X$. Nevertheless, we argue that our choice of $X$ is reasonable as it includes almost all resonances and is not too large as to be dominated by weak interactions that only slightly modify the BBH's orbital parameters \citep{FregeauRasio2007}. However, we miss some of the direct encounters that can lead to a sufficiently large $\Delta e$ to reach $e_{crit}$, while $\Delta E \simeq 0$ \citep{HeggieRasio1996,RandoForastier2024, Trani2024b}.

Another limitation of our approach is with respect to soft binaries. These are usually disrupted in binary-single interactions, but when they undergo BBH-BBH scatterings there is a nonzero chance that they induce a GW merger. Per the limitations of the \cmc code, our analysis does not include soft binary scattering. However, since interactions of soft and hard binaries have a large $\alpha$, their contribution to the merger rate can safely be neglected. However, the interaction among two soft binaries may play a role, as their energy distribution increases very sharply with $a$. For a Saha-like distribution \citep{Heggie1975, Retterer1980}, the creation rate of soft binaries is $C(x)x\propto x^{-7/2}$. If we assume this to roughly equal the interaction rate $\Gamma$, working in the limit of $\alpha=1$ we obtain
\eq{\Gamma(a)\simeq C(x)\abs{\frac{\partial x}{\partial a}}\propto a^{3/2}}
Then, the merger rate is $\mathcal{R}\simeq C(x)p_{merge}\propto a^{11/14}$ (equation~\ref{eq:pIMS}). This points towards the relevance of soft binary interactions, which may explain the enhanced in-cluster merger rate in the $N$-body models of \cite{Barber2024}. In a follow-up study, we consider this possibility in more detail.

\section{Summary and conclusions}
\label{sec:summary}

\subsection{Summary of key findings}
Our work confirms and extends on earlier work on BBH-BBH interactions:
\begin{itemize}
    \item BBH-BBH interactions are less common than BBH-BH interactions (by a factor of $\sim 7$) but, as they are more likely to trigger a GW capture, their contribution to the merger rate is about half that of BBH-BH interactions. This verifies the results of \cite{Zevin2019} for equal-mass BHs.
    \item Stable triple formation is a likely outcome of BBH-BBH interactions. These triples generally have very large semimajor axes, so GW captures due to von~Zeipel-Lidov-Kozai oscillations in dynamically formed stable triples do not significantly contribute to the merger rate. This is in agreement with previous studies, e.g. \cite{Martinez2020}.
\end{itemize}

\noindent Our key novel findings are:
\begin{itemize}
    \item Most BBH-BBH interactions in GCs occur between three-body binaries and not among the primordial binary population. 
    \item The rate and parameters of BBH-BBH interactions in GCs are largely independent of the primordial binary fraction and assumed binary stellar evolution models, as a consequence of the first point. 
    \item BBH-BBH interactions tend to involve the dominant dynamically active BBH that interacts and destroys any further formed softer BBH.
    \item An important consequence of the previous points is that the efficiency of GW captures depends only on the number of BHs in clusters, and is highest in low-mass clusters.
    \item Using a GC population model, we find that BBH-BBH interactions produce mergers at a rate of $\mathcal{R}({z\simeq 0}) = 0.9\units{Gpc^{-3}}\units{yr^{-1}}$, and about $10\%$ of these have $e>0.1$.
    \item This merger rate evolves with redshift according to Fig.~\ref{fig:result:z}. Its peak occurs slightly later than the peak of the cluster formation rate. 
    \item The distribution of eccentricities in GW captures (Fig.~\ref{fig:result:ecc}) is roughly flat in $\log_{10} e$ between $-2.5$ and $0$.
\end{itemize}

\subsection{Enhanced merger rate in low-mass clusters}
The efficiency of GW captures in BBH-BBH interactions scales inversely with the cluster mass, which implies that it is most relevant for less massive clusters and clusters with fewer BHs. Here, we explore the difference between detailed cluster simulations and approximate models that assume a single dynamically active BBH. We argue that the discrepancy at lower masses can be explained by the missing mechanism of GW captures in BBH-BBH interactions.

\begin{figure}
    \centering
    \includegraphics[width=\columnwidth]{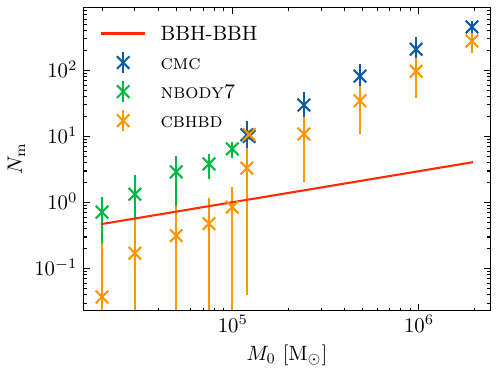}
    \caption{Number of BBH mergers with respect to initial cluster mass for different cluster simulations (crosses with error bars). In red, the number of mergers due to GW captures in BBH-BBH interactions, as in equation~\ref{eq:Nm}. This channel appears to bridge the results of the different codes at lower initial cluster masses.}
    \label{fig:discussion:cBHBd}
\end{figure}

In Fig.~\ref{fig:discussion:cBHBd}, we show the mean number of mergers per cluster in different mass regimes and for different methods of cluster simulation. For the most massive clusters ($\gtrsim 10^5\units{\msun}$) we plot the \cite{Kremer2020} catalogue generated by the \cmc code, whereas for the less massive clusters ($\lesssim 10^5\units{\msun}$) we plot the catalogue of \cite{Banerjee2021b} generated by \textsc{nbody7} \citep{Aarseth2003, Aarseth2012, NitadoriAarseth2012, Hurley2000, Hurley2002}. We assume that these models are detailed enough in their simulation to accurately reproduce the number of GW captures. Overlaid to these, we show the results of resimulating their initial conditions using the fast code \textsc{cBHBd} \citep{Antonini2019, Antonini2023}. We use the same initial conditions and supernova models, and run the simulations for the same amount of physical time.

As can be seen, the number of mergers per initial cluster mass seems to follow a power-law\footnote{This fit is obtained via least-squares regression over all values of $r_h$ in the set of models of \cite{Kremer2020, Banerjee2021b} with equal weighting. If we perform a least-squares fit that allows a power-law dependence on both $M_0$ and $r_h$, we obtain  $N_m \propto M_0^{1.3} r_h^{-0.7}$} $N_m\propto M_0^{1.4}$. The \textsc{cBHBd} code can reproduce the number of mergers at higher cluster masses, but underpredicts the number of mergers at $M_0\lesssim 10^5\units{\msun}$. It appears that adding the effect of GW captures in BBH-BBH interactions bridges the gap with the merger rate of \textsc{nbody7} simulations. We plan to include this effect in \textsc{cBHBd} in the future.

\begin{acknowledgements}
    DMP thanks Miguel Martínez and Giacomo Fragione for help in understanding the \cmc output format. DMP and MG thank Kyle Kremer and Fabio Antonini for discussions. DMP and MG acknowledge financial support from the grants PRE2020-091801, PID2021-125485NB-C22, EUR2020-112157, CEX2019-000918-M funded by MCIN/AEI/10.13039/501100011033 (State Agency for Research of the Spanish Ministry of Science and Innovation) and SGR-2021-01069 (AGAUR). AAT acknowledges support from the Horizon Europe research and innovation programs under the Marie Sk\l{}odowska-Curie grant agreement no. 101103134.
\end{acknowledgements}

\bibliographystyle{aa} 
\bibliography{bibliography} 

\begin{appendix}
\onecolumn
\section{Outcome of the interactions}
\label{app:outcome}
\begin{table*}[htbp]
    \caption{Possible outcomes in a BBH-BBH interaction and the conditions used to classify them in our simulations.}
    \label{tab:outcome}
    \centering
    \begin{tabular}{ll}
    \hline\hline 
    Outcome name & Conditions \\
    \hline
        \multirow{5}{*}{Binary-binary} & The closest pair of BHs is bound (first binary) \\
                                       & The other pair of BHs is bound (second binary) \\
                                       & The two binaries are unbound from each other \\
                                       & The two binaries are moving away from each other \\
                                       & The absolute value of the relative tidal forces is at most $10^{-3}$ times the binding force of the binaries \\ \midrule
        \multirow{5}{*}{Binary-single-single} & The closest pair of BHs is bound (binary) \\
                                       & The other pair of BHs is unbound (singles)\\
                                       & Both singles are unbound from the binary \\
                                       & Both singles are moving away from the binary \\
                                       & The absolute value of the tidal force of each single is at most $10^{-3}$ times the binding force of the binary \\ \midrule
        \multirow{5}{*}{Single-single-single-single} & The closest pair of BHs is unbound \\
                                       & All BHs are moving away from each other \\
                                       & The total energy of the system is positive \\
                                       & The total energy of all subsets of two BHs is positive\\
                                       & The total energy of all subsets of three BHs is positive \\ \midrule
        \multirow{6}{*}{Triple-single} & The closest pair of BHs is bound (inner binary) \\
                                       & One of the other two BHs is bound to the binary (outer BH), whereas the other one is unbound (single) \\
                                       & The single is moving away from the binary \\
                                       & The absolute value of the tidal force of the single is at most $10^{-3}$ times the binding \\& force of the outer component of the triple \\ 
                                       & The triple is \cite{MardlingAarseth2001} stable \\ \midrule
        GW capture & The distance between two BHs is smaller than 10 times the sum of their Schwarzschild radii \\ \midrule
        \multirow{2}{*}{Unresolved} & None of the outcomes above is reached \\
                                    & The simulation time reaches $t=10^6 \sqrt{\frac{(1\units{AU})^3}{G (1 \units{\msun})}}=0.16\,\rm{Myr}$ \\

        \hline
    \end{tabular}
\end{table*}

\begin{multicols}{2}
\section{PN definition of eccentricity}
\label{app:ecc}
As we discussed in Sect.~\ref{ssec:ecc}, in PN theory there are three different values for the eccentricity: temporal $e_t$, radial $e_r$, and angular $e_\phi$, which are only equal in the Newtonian limit. In this Section, we will give an interpretation to these values, following \cite{Memmesheimer2004}. Let us begin by the Kepler equations in Newtonian mechanics
\eq{&R=a(1-e \cos u)\\
&n(t-t_0)=u-e\sin u\\
&\phi-\phi_0=2 \arctan\lrc{\lr{\frac{1+e}{1-e}}^{1/2}\tan\frac{u}{2}}}
where $a$ is the SMA, $e$ is the eccentricity, $u$ is the eccentric anomaly, $n$ is the mean motion, and $t$ is the time. The relative separation between the two bodies is given by the vector $(R \cos \phi, R \sin \phi, 0)$, and the suffix zero indicates the initial conditions. The above equations can be compared to the quasi-Keplerian PN parametrisation of the orbit. For simplicity, in this Section we will only consider 1PN terms, for which the equations read \citep{DamourDeruelle1985}
\eq{&R=a_r(1-e_r \cos u)\\
&n(t-t_0)=u-e_t\sin u\\
&\frac{2\pi}{\Phi}(\phi-\phi_0)=2 \arctan\lrc{\lr{\frac{1+e_\phi}{1-e_\phi}}^{1/2}\tan\frac{u}{2}}}
where $2\pi/\Phi$ is the angular advance of the periastron per orbit. The comparison between the two sets of equations explains the naming of the different eccentricities, as well as giving an intuition on their convergence in the Newtonian limit.

\end{multicols}

\end{appendix}

\end{document}